\documentclass[11pt]{article}
\usepackage{moriond,epsfig}
\usepackage{wrapfig,rotating}
\def\be{\begin{equation}}
\def\ee{\end{equation}}
\def\bea{\begin{eqnarray}}
\def\eea{\end{eqnarray}}

\vspace*{1cm}
\begin{document}

\title{DIFFRACTIVE W AND Z PRODUCTION AT TEVATRON~\footnote{Presented for the CDF collaboration at {\em XLIV$^{th}$ Rencontr\`es de Moriond, QCD and High Energy Interactions, La Thuile,  Aosta Valley, Italy, March 14-21, 2009}.}
}

\author{K. GOULIANOS}

\address{The Rockefeller University, 1230 York Avenue, New York, NY 10065-6399, USA}

\maketitle\abstracts{
Preliminary results on diffractive $W$ and $Z$ production in proton-antiproton collisions at a c.m.s. energy of 1.96~TeV are presented and discussed. Differential cross sections are measured from data collected with the CDF~II detector at the Fermilab Tevatron collider using a Roman pot spectrometer that detects leading antiprotons. The ratio of diffractive to total $W$ production rates (diffractive fraction) agrees with that measured by CDF in Run~I using a rapidity gap diffractive signature. The $W$ and $Z$ diffractive fractions are equal within the measurement uncertainties, which are dominated by a 20\% statistical uncertainty in the diffractive $Z$ measurement. Prospects for extending the diffractive $W$ and $Z$ studies at the Large Hadron Collider are discussed.
}
\section{Introduction}
The CDF collaboration has reported several results on soft and hard diffraction in $\bar pp$ collisions at the Fermilab Tevatron collider using as diffractive signatures large rapidity gaps and/or a leading antiproton (leading protons were not detected). Measured cross sections and certain novel features observed in the data point to a QCD picture of diffraction, in which the exchange is a color-singlet combination of gluons and/or quarks with vacuum quantum numbers, traditionally referred to as the ``Pomeron''~\cite{lathuile07}. 

The most striking observation is a  breakdown of QCD factorization in diffractive dijet, $W$, $b$-quark, and $j/\psi$ production, expressed as a suppression by a factor of $\cal{O}$(10) of the production cross section relative to theoretical expectations. Of equal importance is the finding of a breakdown of Regge factorization in soft diffraction by a similar factor~\cite{lathuile07}. Combined, these two results strongly support a hypothesis that the factorization breakdown is due to a saturation of the probability of forming a rapidity gap by an exchange of a color-neutral construct of the underlying parton distribution function (PDF) of the proton. Renormalizing the ``gap probability'' to unity over all $(\xi,t)$ phase space, where $\xi$ is the forward momentum loss fraction of the leading (anti)proton, corrects for the unphysical effect of overlapping diffractive rapidity gaps and leads to an agreement between theory and experiment~\cite{lathuile07}. The gap probability renormalization model is further supported by soft-diffraction CDF results on double-diffraction (central gap), multi-gap diffraction (double-gap to single-gap ratios are non-suppressed), energy dependence of the total single-diffractive cross section, $\sigma^D_{tot}\rightarrow$~constant as $s\rightarrow\infty$, and a relationship between the intercept and slope of the Pomeron trajectory~\cite{blois07}.

\begin{figure}[ht]
\centerline{\includegraphics[width=0.75\textwidth]{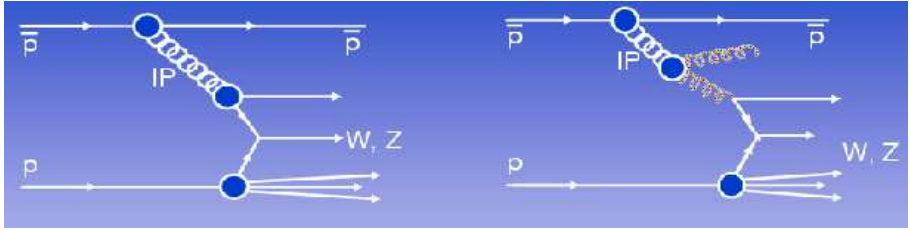}}
\caption{Diffractive $W/Z$ production diagrams.}\label{wz_diagrams}
\vspace*{-1em}
\end{figure}

The $\cal{O}$(10) suppression of diffractive dijet production at the Tevatron is based on the proton PDF extracted from diffractive deep inelastic scattering (DDIS) at the DESY $ep$ Collider HERA~\cite{lathuile07}. While  
 no DDIS suppression is expected in certain models, e.g.~\cite{collins}, the primary exchange in DDIS is a $q\bar q$ pair (Fig.~\ref{wz_diagrams}, left), while production by a gluon is suppressed by a factor of $\alpha_s$. The latter can be distinguished from quark production by an associated jet (Fig.~\ref{wz_diagrams}, right)~\cite{cdf_W}. In contrast to $W/Z$, diffractive dijets are mainly produced by a $gg$ exchange. The dijet rates at the Tevaytron are calculated using a gluon PDF extracted from DDIS. A more direct comparison between diffraction at the Tevatron and at HERA can be made by measuring diffractive $W$ production at the Tevatron, which is dominated by a $q\bar q$ exchange. In Run~I, only the overall diffractive $W$ fraction was measured~\cite{cdf_W}. In Run~II, CDF measures both the $W$ and $Z$ diffractive fractions, and has developed a method that completely determines the $W$ kinematics and can be used to measure the $x_{Bj}$ ($x$-Bjorken) dependence of the diffractive structure function (DSF). 
 
\noindent Figure~\ref{wz_diagrams} shows schematic Feynman diagrams for diffractive $W/Z$ production. In leading order, the $W/Z$ is produced by a quark in the Pomeron (left); production by a gluon is suppressed by a factor of $\alpha_s$ and can be distinguished from quark production by an associated jet~\cite{cdf_W} (right). 

\section{Detector}
\begin{figure}[ht]
\vspace*{-1em}
\centerline{\includegraphics[width=0.8\textwidth]{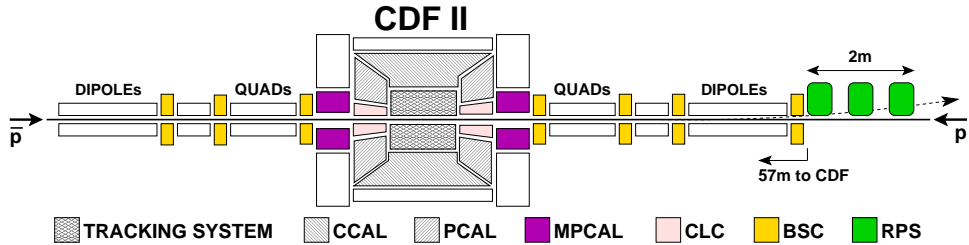}}
\caption{The CDF II detector.}\label{Fig:detector}
\end{figure}

The CDF~II detector is shown schematically in Fig.~\ref{Fig:detector}~\cite{excl2j}. 
The components of the main detector~\cite{cdf2detector} used in the diffractive program are the tracking system, the central (CCAL), plug (PCAL), and forward (FCAL) calorimeters, and the \v{C}erenkov luminosity counters (CLC). 
The diffractive program benefited from dedicated triggers and a system of special forward detectors.
The following forward detectors were employed~\cite{excl2j}:   
\vspace*{-0.8em}
\begin{itemize}
\addtolength{\itemsep}{-0.85em}
\item  RPS (Roman Pot Spectrometer)~-~detects leading $\bar p$'s at $\sim 0.03<\xi\equiv 1-p_{||}<0.09$;
\item MPCAL (MiniPlug Calorimeters)~-~measure $E_T$ and $(\theta,\phi)$ at $\sim 3.5<|\eta|<5.5$;
\item BSC (Beam Shower Counters)~-~identify rapidity gaps at $\sim 5.5<|\eta|<7.5$.
\end{itemize}

\section{Diffractive W and Z Measurement}
The data analysis is based on events with RPS tracking from a data sample of approximately $0.6$~fb$^{-1}$. In addition to the $W/Z$ selection requirements (see below), a hit in the RPS trigger counters and a RPS reconstructed track with $0.03<\xi<0.1$ and $|t|<1$ are required. 
A novel feature of the analysis is the determination of the full kinematics of the $W\rightarrow e\nu/\mu\nu$ decay, which is made possible  by obtaining the neutrino $E_T^\nu$ from the missing $E_T$, as usual, and $\eta_\nu$ from the formula $\xi^{\rm RPS}-\xi^{\rm cal}=(E_T/\sqrt{s})\exp[-\eta_\nu]$ , where $\xi^{\rm cal}=\sum_{\rm i(towers)}(E_T^i/\sqrt{s})\exp[-\eta_i]$. 
\hspace*{-1em}\includegraphics[width=0.5\textwidth]{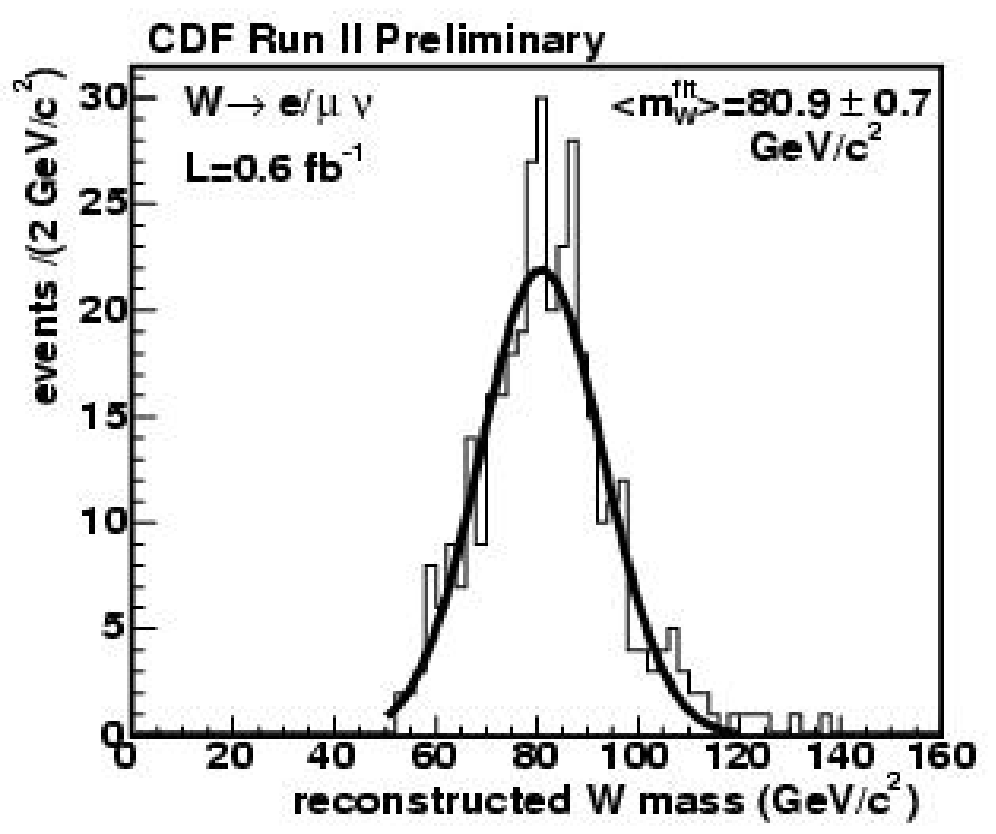}
\begin{figure}[h]
\end{figure}
\setcounter{figure}{3}
\vglue -19em
\hspace{20em}\begin{minipage}{20em}
The CDF $W/Z$ selection requirements are $E_T^{e,\mu}>25$~GeV, $40<M_T^W<120$~GeV, $66<M^Z<116$~GeV, and vertex $z$-coordinate $z_{vtx}<60$~cm. The $W$ mass distribution for events with $\xi^{\rm CAL}<\xi^{RPS}$ is shown in Fig.~3 along with a Gaussian fit. The obtained value of $M_W^{\rm exp}=80.9\pm 0.7$~GeV is in good agreement with the world average $W$-mass of $M_W^{\rm PDG}=80.403\pm 0.029$~GeV~\cite{PDG}. 
Figure~\ref{Fig:xi_cal} shows the $\xi^{\rm CAL}$ distributions of the $W/Z$ events satisfying different selection requirements. 
In the $W$ case, the requirement of $\xi^{\rm RP}>\xi^{\rm CAL}$ is very effective in removing the overlap events in the region of $\xi^{\rm CAL}<0.1$.
\end{minipage}

\small{ Fig. 3. Extracted $W$ mass and Gaussian fit.}
\vglue -1em
\begin{figure}[ht]
\centerline{\includegraphics[width=0.5\textwidth]{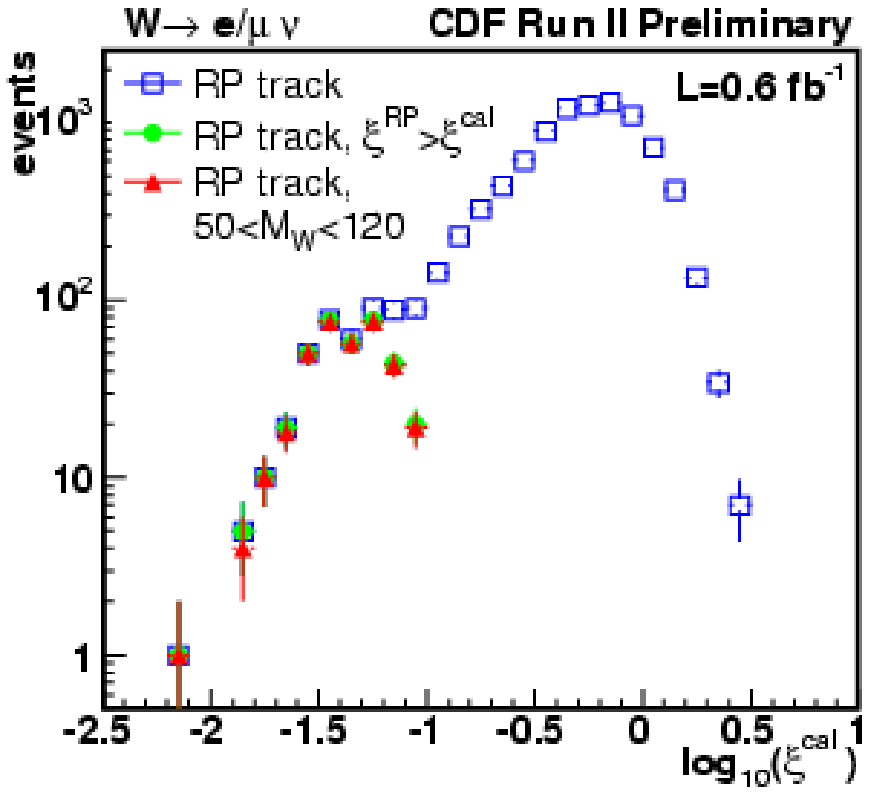}
\includegraphics[width=0.52\textwidth]{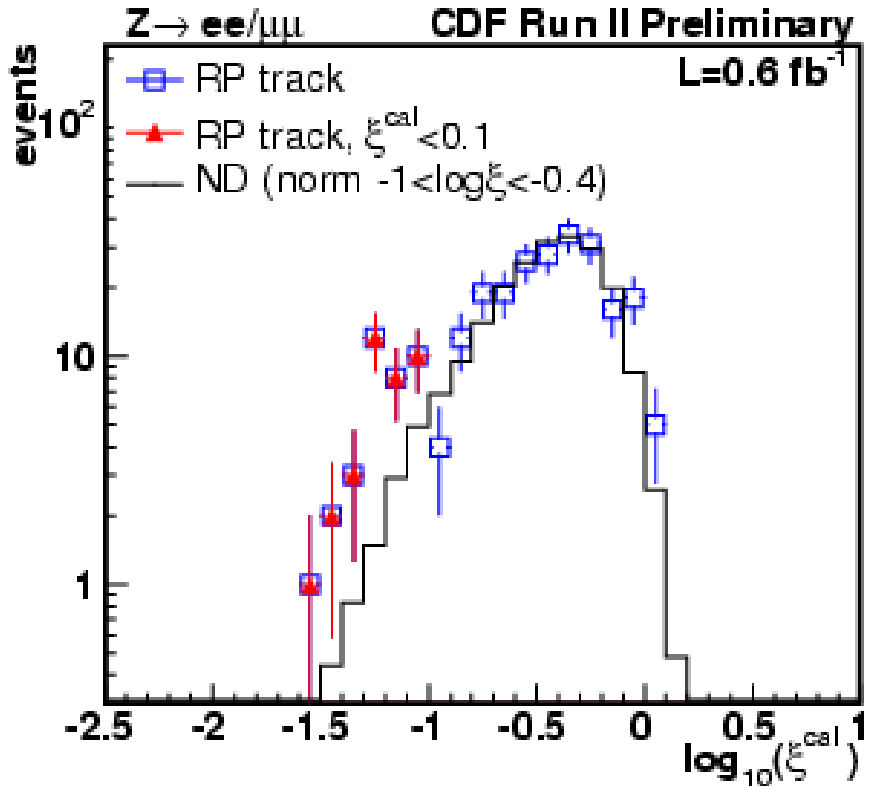}}
\vspace*{-1em}
\caption{The $\xi^{\rm CAL}$ distribution for various $W$ (left) and $Z$ (right) event samples.}
\label{Fig:xi_cal}
\end{figure}
A mass cut of $50<M_W<120$~GeV has the same effect. In the $Z$ case, the $\xi^{\rm CAL}$ distribution of all $Z$ events is used and is normalized to the RP-track distribution in the region of $-1<\log\xi^{\rm CAL}<-0.4$ ($0.1<\xi^{\rm CAL}<0.4$) to obtain the ND background in the diffractive region of $\xi^{\rm CAL}<0.1$.  
 
Taking into account the RPS acceptance of $A_{\rm RPS}\approx 80$~\%, the trigger counter efficiency of $\epsilon_{\rm RPStrig}\approx 75$~\%, the track reconstruction efficiency of $\epsilon_{\rm RPStrk}\approx 87$~\%, and then multiplying by 2 to include the production by $\bar pp\rightarrow X+W/Z+p$ and correcting the number of ND events for the effect of overlaps due to multiple interactions by multiplying by a factor of $f_{\rm 1-int}\approx 25$~\%, the diffractive fractions are obtained as
$R_{W/Z}=2\cdot N_{SD}/A_{\rm RPS}/\epsilon_{\rm RPStrig}/\epsilon_{\rm RPStrk}/(N_{\rm ND}\cdot f_{\rm 1-int})$, which yiel the results:\\  
\vglue -0.75em    
$R_W(0.03<\xi<0.10,\,|t|<0.1)=[0.97\pm 0.05\;\mbox{(stat)}\pm 0.11\;\mbox{(syst)]}\%,$

$R_Z(0.03<\xi<0.10,\,|t|<0.1)=[0.85\pm 0.20\;\mbox{(stat)}\pm 0.11\;\mbox{(syst)]}\%.$

\noindent The $R_W$ value is consistent with the Run~I result of:

$R_W(0.03<\xi<0.10,\,|t|<0.1)=[0.97\pm 0.47]~\%\;(Run~I)$, 

\noindent obtained from the measured value of $R^W(\xi<0.1)=[0.15\pm0.51\;\mbox{(stat)}\pm0.20\;\mbox{(syst)}]\%$~\cite{cdf_W} multiplied by a factor of 0.85 tto account for the reduced ($\xi$-$t$) range in Run~II.

\section{Prospects for the LHC}
Diffractive $W$ and $Z$ production at the LHC can provide information in the following areas:
\vspace*{-0.5em}
\begin{itemize} 
\addtolength{\itemsep}{-0.5em} 
\item energy dependence of the production cross sections;
\item $x$-Bjorken dependence of the diffractive structure function in $W$ production;
\item exclusive $Z$ production.
\end{itemize}
The energy dependence of the cross sections is crucial for understanding the nature of the unitarity constrains that lead to the observed breakdown of factorization; the $x$-Bjorken distribution in $W$ production, the measurement of which requires measuring the momentum of the leading diffracted proton in Roman pot detectors, can be used to differentiate among various models of the Pomeron; and exclusive $Z$ production, which can be used to search for beyond the standard model theories of the Pomeron~\cite{alanwhite}.

\section{Summary and Conclusion}
Preliminary results on diffractive $W$ and $Z$ production obtained by the CDF~II collaboration from proton-antiproton collisions at a c.m.s. energy of 1.96~TeV have been presented. Differential cross sections were  measured from data triggered by a leading antiprotons detected in a Roman pot spectrometer in association with a central high transverse momentum electron or positron. The obtained ratio of diffractive to total $W$ production rates (diffractive fraction) agrees with that measured by CDF in Run~I where diffraction was identified by a large rapidity gap signature. The $W$ and $Z$ diffractive fractions are equal within a $\sim$20\% measurement uncertainty dominated by the statistical uncertainty in the diffractive $Z$ measurement. A method of completely determining the diffractive $W$ kinematics by taking advantage of the momentum loss measurement of the leading antiproton using the RPS has been disussed. This method can be used for measuring the $x$-Bjorken dependence of the diffractive structure function in $W$ production at the Large Hadron Collider over a $x$-Bjorken range. Comparison of the $W$ and dijet $x$-Bjorken distributions is critical to understanding the partonic nature of the colorless diffractive exchange.

\section*{Acknowledgments}
Many thanks to my colleagues at The Rockefeller University and in the CDF collaboration, whose contributions to the diffractive program have opened new windows to the intriguing world of diffraction and exclusive production and made this work possible.


\section*{References}
\begin{thebibliography}{99}
\bibitem{lathuile07}K.~Goulianos, ``Diffraction and Exclusive (Higgs?) Production from CDF to LHC''? in {\em Les Rencontres de Physique de la Vallee d'Aoste: Results and Perspectives in Particle Physics}, La Thuile, Italy, 4-10 March 2007, pp. 177-200, Frascati Physics Series, Special 44 Issuee, editor Mario Greco.
%
\bibitem{blois07}K.~Goulisnos, ``Diffraction at CDF,'' in {\em 12$^{th}$ International Conference on Elastic and Diffractive Scattering (Blois Workshop) - Forward Physics and QCD}, DESY, Hambyrg, Germany 21-25 May 2007, arXiv:0712.3633V2[hep-ph] 5 Jun 2008, pp.~137-144; ib. ``{Pomeron intercept and slope: the QCD connection}, pp.~248-253; submitted to Phys. Lett. {\bf B}, arXiv:0812.4464v2 [hep-ph].
%
\bibitem{collins}J. Collins,
J. Phys. G {\bf 28}, 1069 (2002); arXiv:hep-ph/0107252. 
%
\bibitem{excl2j}T.~Aaltonen {\it et al.} (CDF Collaboration), 
Phys. Rev. D {\bf 77}, 052004 (2008).
%
\bibitem{cdf2detector}D.~Acosta {\it et al.}  (CDF Collaboration), Phys.\ Rev.\ D {\bf 71}, 032001 (2005).
%
\bibitem{cdf_W} Abe et al. (CDF Collaboration), 
Phys. Rev. Lett. {\bf 78}, 2698-2703 (1997).
%
\bibitem{PDG}W.-M. Yao et al., Journ. Phys. G {\bf 33}, 1 (2006)
and 2007 partial update for 2008, http://pdg.lbl.gov.
%
%
%
\bibitem{alanwhite}A.~R.~White, Phys. Rev. D{\bf 72}, 036007 (2005).
\end{thebibliography}
\end{document}